
\PHYSREV
\unnumberedchapters
\date={November 1993}
\Pubnum={\caps UPR-593-T}
\titlepage
\title{Cosmic Censorship Violation for a Class of Supersymmetric Solitons}
\frontpageskip=0.5\medskipamount plus 0.5 fil
\author{ Mirjam Cveti\v c and Donam Youm}
\address{Department of Physics\break
University of Pennsylvania\break
Philadelphia, PA 19104--6396\break}

\abstract{We study vacuum domain walls in a class of four-dimensional
$N=1$ supergravity  theories where along with the  matter field, forming  the
wall, there is  more than one  ``dilaton'', each respecting
$SU(1,1)$ symmetry in their sub-sector. We find {\it supersymmetric} (planar,
static)  walls,  interpolaing between  Minkowski  vacuum  and a new class of
supersymmetric vacua which  have  a naked  (planar) singularity.
Such walls    provide the first example of supersymmetric classical
``solitons'' with naked singularities,  and thus violate the
(strong) cosmic censorship conjecture within a supersymmetric theory.}

\line{PACS \# 11.17+y, 04.20.-q, 04.65.+e, 10.30.Pb \hfill}
\endpage

\REF\P{R. Penrose, Rev. Nuovo Cimento {\bf\ 1}, 252 (1969).}

\REF\FOOTBB {{\it E.g.}, see a recent work  addressing formation of naked
singularities   in  the collision of black holes in the
space-time with the positive cosmological constant:
D. R. Brill, G. T. Horowitz, D. Kastor and J. Traschen, {\it Testing Cosmic
Censorship with Black Hole Collisions}, NSF-ITP-93-78, UMHEP-387, gr-qc/
930714, and
J. Horne and G. T. Horowitz, {\it Cosmic Censorship and the Dilaton},
NSF-ITP-93-95, YCTP-P17-93, hep-th/9307177.}

\REF\GA {G. Gibbons, Nucl. Phys. {\bf B207}, 337 (1982).}

\REF\GB {G. Gibbons,
in {\it Supersymmetry, Supergravity and Related Topics}, F. del Aguila
{\it et al}. eds. (World Scientific, Singapore, 1985), {\it p} 124-166.}

\REF\KLOPP{R. Kallosh, A. Linde, T. Ort\' in, A. Peet and A. Van Proeyen,
Phys. Rev. {\bf D 46}, 5278 (1992).}
\REF\FOOTAA{Since the space-time is Minkowski only on one side of
the wall, it does
not comply with the weak cosmic censorship conjecture.}
 \REF\FOOTCC{For black hole configurations, a counter-example with
asymptotically flat space-time has been
found by R. Kallosh  (private
communications). This example  corresponds to a Kerr-Newman
black hole with $M = Q$,
 with the rotation parameter $a$  passing its
extremal limit.  Here, $M$ and $Q$ are the mass and the charge of the
black hole, respectively.  Such a black hole is
supersymmetric, however it has a naked singularity.  There is also an example
of extreme charged black holes in $5d$ Kaluza-Klein theories with naked
singularities,
found in Ref. \GA\ .}
 \REF\FOOTCP{ L. Romans, Nucl.
Phys. {\bf B383}, 395 (1992). Here
supersymmetric black holes with a  cosmological constant were found to have
naked singularities.  In this case the asymptotic
space-time is not flat. }

Some of the solutions of gravity theory  correspond to  configurations
with naked space-time singularities, $i.e.$, singularities which are
not hidden behind horizons.  This uncomfortable feature is remedied by
Penrose's conjecture which states that generic initial conditions
do not evolve
to form naked singularities.\refmark\P\  Such a conjecture is difficult to
prove and dynamical formation of naked  singularities has been addressed only
for specific cases.\refmark{\FOOTBB}

On the other hand, it has
been observed \refmark{\GA, \GB, \KLOPP}\ that in supersymmetric theories
the allowed  black hole configurations are
only those with mass $M$ bounded from below by the Bogomol'nyi bound,
{\it e.g.}, $M \ge \sqrt{P^2 + Q^2}$, where $P$ and $Q$ correspond to
the magnetic and electric charges of the black hole, respectively.
Incidentally, such a bound coincides with the one of cosmic censorship. Namely,
black holes in supersymmetric theories have singularities hidden behind
(or at) the
horizon.  This observation prompted a conjecture \refmark{\KLOPP}\
that { supersymmetry acts  as a cosmic censor}, {\i.e.},
(solitonic) configurations in supersymmetric theories do not have naked
singularities.   The conjecture applies\refmark{\KLOPP}\
only to the configurations   which have asymptotically flat space-time, {\it
i.e.}, they comply with the weak cosmic censorship conjecture.
It has been proven\refmark{\GB}\ in a related context
that in ungauged  extended supergravity theories there are no  classical
  ``solitons'' without horizons.
Such solitons were assumed to have a non-trivial structure in the interior, and
to tend at large distances toward a supersymmetric vacuum without matter
sources.
\REF\CGR {M. Cveti\v c, S. Griffies and S.-J. Rey, Nucl. Phys. {\bf B381},
301 (1992).}
\REF\CG {M. Cveti\v c and S. Griffies, Phys. Lett. {\bf B285}, 27 (1992).}
\REF\CDGS {M. Cveti\v c, R. L. Davis, S. Griffies and H. H. Soleng,

Phys. Rev. Lett. {\bf 70}, 1191 (1993).}
\REF\GIBBIII{G. Gibbons, Nucl. Phys. {\bf 394}, 3 (1993).}
\REF\C {M. Cveti\v c, Phys. Rev. Lett. {\bf 71}, 815 (1993).}

In this letter we present supersymmetric vacuum domain wall configurations
 in a class of four-dimensional ($4d$) $N=1$ supergravity
models, where along with the matter field, forming the wall, there are $n\ge
2$ ``dilaton'' fields,  each  respecting $SU(1,1)$ symmetry in their
sub-sector.  Such  walls are {\it supersymmetric}
(planar, static)   configurations, where on one side of the wall the
space-time is flat, while on the other side the space-time has a (planar)
naked singularity. They therefore provide an example of classical
supersymmetric solitons, which   violate the (strong) cosmic
censorship.\refmark{\FOOTAA, \FOOTCC, \FOOTCP}\
One is able to trace the origin of these singularities to the
fact that these walls interpolate between
(supersymmetric) Minkowski  vacuum, {\it i.e} the flat space-time   and a {\it
new
class of supersymmetric vacua}  associated  with a non-trivial
source of the dilatons.  Namely,  in such supersymmetric vacua
$n\ge 2$ dilatons  render
the vacuum energy
positive and the stress-energy tensor violates the strong energy condition.
Such vacua are in sharp contrast with
the   Minkowski and anti-deSitter
space-times, {\it i.e.}, unique supersymmetric vacua without matter sources.
The latter ones
were used in the proof  of  ``no solitons without horizons'' in the
ungauged extended supergravity.\refmark\GB\
 Such walls should also  be contrasted with
ordinary ($n=0$)\refmark{\CGR -\GIBBIII} and
dilatonic ($n=1$)\refmark{\C}\  supergravity  walls.

\REF\FOOTO{An alternative  description can be in terms of linear
super-multiplets. Since our solutions  have imaginary components $a_i$
set to zero  and  gauge fields are turned off, both descriptions
are equivalent [{\it
E.g.}, P. Adamietz, P. Binetruy,  G.Girardi and R. Grimm, Nucl. Phys.
{\bf B401}, 257
(1993) and references therein.].  Note, that linear multiplets have no
superpotential.}
\REF\FOOTP{ K\" ahler  potential
$K(S_i,\overline {S}_i) = -\sum^n _{i=1}\beta_i\ln(S_i + \overline
{S}_i) $  with $\beta_i>0$ yields a  straightforward generalization.
We motivate the choice $\beta_i=1$ from  examples in
string theory and no-scale supergravity models.}
\REF\FKN{ {\it E.g.}, J. Ellis, C. Kounnas and D. Nanopoulos, Nucl. Phys.
{\bf 247B}, 373 (1984) and references therein.}
\REF\WITTEN{E. Witten, Phys. Lett. {\bf 155B},  151 (1984).}
\REF\FOOTQ{ G. Lopes Cardoso and B.
Ovrut, {\it Supersymmetric Calculation of Mixed K\" ahler-Gauge  and Mixed
K\" ahler-Lorentz Anomalies},
CERN-TH.6961/93, hep-th/9308066, and references therein. }

We consider  $4d$ $N = 1$ supergravity theory with $n\ge 2$ dilatons $S_i
\equiv {\rm e}^{-2\phi _i} + ia_i$
($i = 1, ..., n$), which  we choose to describe  as  scalar components
of the chiral super-multiplets.\refmark\FOOTO\   Dilatons have  K\" ahler
potential $K(S_i,\overline {S}_i) = -\sum^n _{i=1}{\rm ln}(S_i + \overline
{S}_i) $  and no superpotential ($W(S_i)\break = 0$), thus respecting $SU(1,1)$
non-compact symmetry  in each sub-sector.\refmark\FOOTP\
The scalar component $T$
of the matter multiplet has  K\" ahler potential
$K_M (T, \overline{T})$ and superpotential $W_M(T)$, which allow for isolated
minima of the matter potential, and thus for $T$ to form the wall.
A crucial property of the effective action is that it can be written in terms
of the separable K\" ahler potential
$K = K_M (T, \overline{T}) + K(S_i , \overline{S}_i )$ and superpotential
  $W=W_M(T)$, which  depends {\it only} on the matter field $T$.

Such a class of  supergravity theories is motivated by the
no-scale supergravity models\refmark\FKN\ as well as by the
effective theory of  $4d$
superstring vacua.\refmark\WITTEN\  In the latter case   one field
corresponds to the dilaton field of the string theory
and  the other $(n-1)$   fields are  the  compactification moduli.
Note, however,  that for  superstring vacua the $n-1$  moduli cannot be
rewritten as scalar components of linear multiplets and
  matter fields in general
do  couple to the corresponding moduli fields
in the K\" ahler potential.  The above proposed class of supergravity
models should thus be viewed primarily as  a specific framework  which
illustrates the existence  of classical supersymmetric solitons with naked
singularities.

The scalar part of the tree level $N = 1$ supergravity
Lagrangian is of the form:
$$
{\cal {L}} =\sqrt{-g} [- {1\over 2} R + K_{T \overline{T}} g^{\mu \nu}
\partial _\mu T \partial _\nu \overline{T} + \sum^n _{i=1}
g^{\mu \nu} \partial_\mu \phi_i \partial_\nu \phi_i -
2^{-n}{\rm exp}(2\sum^n_{i=1}\phi_i)\tilde V]\eqn\1
$$
where
$$
\tilde V = {\rm e}^{K_M} (K^{T \overline{T}} |D_T W_M|^2 -
(3-n)|W_M|^2 )\eqn\2
$$
is the part of the potential, which  depends only on the matter field $T$.
Here $K^{T \overline{T}} \equiv (K_{T \overline
{T}})^{-1} \equiv (\partial_T \partial_{\overline {T}}K)^{-1}$ and $D_T
W_M \equiv {\rm e}^{-K_M}\partial_{T}({\rm e}^{K_M} W_M)$.  We use
the convention  $(+---)$ for the metric and the gravitational
constant $\kappa = 8\pi G=1$.
In \1\ we have already set $a_i = 0$ (which turns out to correspond to the
solution of equation of motion) and turned off the gauge fields. We also assume
that the  models are free of  mixed K\"ahler-Lorentz
 anomalies.\refmark\FOOTQ\

We assume that $\tilde V$  has isolated
minima, thus allowing for $T$ field to form a wall configuration.
Note, that $\tilde V$ is
modified due to the presence
of $n$  real scalar fields $\phi_i$, which yield
an additional contribution ${\rm e}^K K^{S_i \overline {S}_i} |D_{S_i}W|^2
= {1 \over 2} {\rm e}^{2\phi_i}{\rm e}^{K_M}|W_M|^2$ to $\tilde{V}$ for each
field $\phi_i$.   For  supersymmetric minima, $D_{T} W_{M} = 0$ and
$\tilde{V} = (n - 3){\rm e}^{K_M} |W_M|^2$.  Therefore, at such  minima
dilatons  $\phi_i$  screen the matter potential by
$2^{-n}\prod^n_{i=1}{\rm e}^{2\phi_i}$  (see Eq.(1)) as well as changing
an overall
scale factor of  the matter potential from $-3$ (for the ordinary
supersymmetric
vacuum) to $(-3 + n)$, thus  rendering the matter potential (2)
less negative. For
 supersymmetric  minima $\tilde V$ is  non-positive for $0\le n\le 2$,
it vanishes identically for $n = 3$,  and it is  always non-negative for $n
\geq 4$.
We have therefore constructed examples of supergravity models, where
supersymmetric minima can have {\it positive} vacuum energy.
This is counter to the prevailing lore that for the
supersymmetric  vacua the vacuum energy  is non-positive.

\REF\FOOTK{One can obtain explicit numerical solutions of Eqs.\5\
for  a wall of any thickness with similar qualitative features  as discussed
for
the thin walls in the text. Namely, on one side of the wall the matter field
$T(z)$ reaches the
supersymmetric minimum at a finite   $z_{sing}$. At
$z_{sing}$ the conformal factor $A(z)$ degenerates, yielding the space-time
 singular.}

\REF\FOOTF {Outside the thin wall ($\partial_z T=0$)
 Einstein's equations for the conformal factor $A(z)$ and Euler-Lagrange
equations for
$\phi _i (z) = \phi (z) + (\phi _i )_0$ ($i = 1,...,n$), as derived from
Lagrangian (1),
 are of the form:$$
\eqalign{-H^{\prime} - {1 \over 4}H^2 &= n(\phi^\prime )^2 +
(n-3)\alpha_i^2 {\rm e}^{2n\phi}A \cr
{3 \over 4}H^2 &= n(\phi^\prime )^2 - (n-3)\alpha_i^2 {\rm e}^{2n \phi}A \cr
 {(A\phi^\prime )}^\prime &= (n-3)\alpha_i^2{\rm e}^{2n \phi}A^2 \cr}
$$where $H=A^\prime/A$ and $A^\prime\equiv \partial_zA$.
Here
$\alpha_i$ ($i=1,2$ for $z>0$ and $z<0$, respectively) is defined
after Eq.(6) in terms of the matter K\" ahler potential and superpotential
on either side of the wall.  Note, that  $\tilde V_i =(n-3)\alpha_i^2$ (see
Eq.(2)) corresponds
to the value of the matter part of the potential  on either side of the wall.
[Some of the solutions of the the above equations have  been found by H.
Soleng (unpublished).]
The stress-energy tensor at
the wall is of the form $T_{\mu \nu } = \sigma \delta (z)
(1, -1, -1, 0)$.   The matching conditions
 [W. Israel, Il Nuovo Cimento {\bf 44B}, 1 (1966); Erratum {\it
ibid} {\bf 48B}, 463 (1967).]
 for $A (z)$ and
$\phi (z)$  at the wall  are:
$ A^\prime(0^+) - A^\prime(0^-) =
-{\sigma}$ and  $\phi^\prime (0^+) - \phi^\prime (0^-) =
{1 \over 2}{\sigma}.$  They  impose boundary conditions on
solutions of the above Einstein's and Euler-Lagrange equations.
Note, that such matching conditions  are  automatically satisfied by the
 solutions of  Eqs.\5\  .}
\REF\FOOTFF{The results obtained from Lagrangian (1) should be contrasted with
 those for the   Jordan-Brans-Dicke ($JBD$)  Lagrangian which  in the
Einstein  frame is of the form:
$${\cal {L}}_{JBD} = \sqrt {-g} \left [-{1\over 2}R+{1\over 2}g^{\mu \nu}
\partial_{\mu}\tilde \phi \partial _{\nu} \tilde \phi + {\rm e}^
{-\beta \tilde\phi}{\cal {K}} - {\rm e}^{-2 \beta \tilde\phi } V  \right],
$$
where $\beta \equiv (\omega + {3 \over 2} )^{-{1 \over 2}}$, while
 ${\cal {K}}$ and $V$ correspond to the kinetic  and  potential
energy of the matter field forming the wall, respectively.
 $\omega$    is the coefficient in  front of the kinetic energy of the
JBD field
$\Phi\equiv {\hbox{e}}^{\beta
\tilde \phi}$  in the JBD frame defined by
${g_{JBD}}_{\mu\nu}=g_{\mu\nu}/\Phi$ [P. Jordan , Z. Phys. {\bf 157}, 112
(1959); C.  Brans and C. Dicke Phys. Rev. {\bf 124}, 925 (1961).] .
Outside the thin wall   ${\cal {K}}=0$.
In this case  the  Lagrangian (1)  can be written in the form of the
$JBD$ Lagrangian  provided
 $\phi _i (z) = \tilde\phi (z)/\sqrt{2n} + (\phi _i )_0$ ($i = 1,...,n$),
  $V= 2^{-n}{\rm e}^{2\sum^n_{i=1} (\phi _i)_0} \tilde{V}$, and  $\omega
\equiv {2\over n} - {3 \over 2}$. However, inside the wall region, ${\cal
{K}}\ne 0$, and
the $JBD$  Lagrangian  {\it is  not}
 of the form (1). Thus, for the $JBD$ Lagrangian the
 matching conditions  at the wall  are different:
$A^{\prime}(0^+ ) - A^{\prime }(0^- ) = -\sigma$,  and $\tilde \phi ^{\prime}
(0^+ ) - \tilde\phi^{\prime} (0^- ) = -{3\over 2} \beta \sigma$ and are
obviously not satisfied by the solutions of Eqs.\5\ .
}

\REF\FOOTL{ Type $II$ and type $III$ walls correspond to the walls with
$\alpha_{1}
\ne 0$  and  $\alpha_{2} \ne 0$, {\it i.e.}, on both sides of the wall
$W_M\ne 0$.
 Type $II$ walls  have $W_{M}$ traversing zero,  $+$ sign in Eq.\6\  and
conformal factors $A(z)_{1,2} = [1 - (n - 1)\alpha _{1,2}|z|
]^{2 \over n-1}$, {\it i.e.}, they have naked singularities on both sides
of the wall.   Type $III$ walls have  $W_{M} \ne 0$
everywhere, $-$ sign in  Eq.\6\  and conformal factors
$A(z)_{1} = [1 -
(n - 1)\alpha _{1}|z|]^{2 \over n-1}$ for $z<0$ and
$A_{E}(z)_{2} = [1 + (n - 1)\alpha_{2}|z|]^{2
\over n-1}$ for $z>0$, {\it i.e}, they  have
a naked singularity on one side of the wall ($z<0$) and  are
geodesically complete on the  other side of the wall ($z>0$).}

\REF\FOOTH {Note that ordinary
 supergravity
 walls ($n=0$)   have the thin wall solution  $A(z) = [1 - \zeta \alpha_i
z]^{-2}$ (i=1,2),  while dilatonic walls ($n=1$)  have the
thin wall solution $A(z) = {\rm exp} [2\zeta
\alpha_i z]$ (i=1,2).
 On the side with larger $\alpha_i$, the space--time of the
ordinary walls\refmark{
\CDGS ,\GIBBIII}  has no singularities with $|z|=\infty$ corresponding
to the Cauchy horizon, while  for dilatonic walls  the singularity
at $|z|=\infty$ coincides
with  the horizon.\refmark{\C }\ Note that
on that  side of the  ordinary
 and dilatonic supergravity  walls  the weak and the
dominant energy conditions are violated, while the strong energy condition
is satisfied.}

\REF\FOOTI {For the null geodesics the affine parameter
$\lambda_{0} (z) = E^{-1} \int ^z A(z^\prime )dz^\prime$ is
finite as $|z| \rightarrow {1 \over (n - 1)}\alpha_1$.   Here  $E$ is a
conserved energy parameter.
The proper time of the massive test  particle is
$\tau (z) = \int ^z {{mA(z^\prime )dz^\prime } \over
{\sqrt {E^2 - m^2 A(z^\prime )}}}$.  With $A(z) = [1 - (n - 1)
\alpha_1 |z|]^{2 \over (n-1)}$ ($n \geq 2$), $\tau (z)$ is finite as $|z|
\rightarrow {1 \over n-1}\alpha_1$.}

\REF\FOOTM{It can be shown, by applying Killing spinor identities [R. Kallosh
and T. Ort\' in, {\it Killing Spinor Identities},
SU-ITP-93-16, hep-th/9306085.]
that our bosonic solutions do not acquire  quantum corrections,
which respect $N=1$ supersymmetry.  }

\REF\DK {A. D. Dolgov and I. B. Khriplovich, Gen. Rel. Grav. {\bf 21},
13 (1989).}

We start with the metric
{\it Ansatz} for planar (in ($x,y$) plane), static domain wall
solutions:
$$
ds^2 = A(z)(dt^2 - dz^2 - dx^2 - dy^2),\eqn\3
$$
and  the scalar fields $T(z)$, and $\phi_{i} (z)$ depend only on $z$.  Using a
technique of the generalized Israel-Nester-Witten  form developed in
Ref. \CGR\ for the
study of supergravity walls, one obtains the following  Bogomol'nyi bound
for the
energy density $\sigma$  of the planar domain wall configuration:
$$
\sigma - |C| = \int^{\infty}_{-\infty} [ -\delta _{\varepsilon}
\psi^{+} _{i}g^{ij} \delta _{\varepsilon} \psi _{j} + K_{T \overline {T}}
\delta _{\varepsilon}\chi ^+ \delta _{\varepsilon} \chi + \sum ^n_{j=1} K_{S_j
\overline{S} _j} \delta _{\varepsilon}\eta^+_j \delta _{\varepsilon} \eta _j
]dz \geq 0. \eqn\4
$$
This bound is saturated iff the supersymmetry variations $\delta _{\varepsilon}
\psi _{\mu}$, $\delta _{\varepsilon}\chi$, and $\delta _{\varepsilon} \eta _j$
of the fermionic partners of the fields $g_{\mu \nu}$, $T$ and $S_j$,
respectively, vanish.  For this case, one has {\it supersymmetric} bosonic
backgrounds, and the metric and scalar fields satisfy coupled first order
differential equations (self-dual or Bogomol'nyi equations):
$$
\eqalign{0&={\rm Im} (\partial _{z}T {{D_{T}W_{M}} \over {W_M}})
\cr \partial_{z} T &= -\zeta (2^{-n}A{\rm e}^{2\sum ^n_{1}\phi _i} )^{1/2}
{\rm e}^{K_M /2} |W_M| K^{T \overline{T}} _M
{{D_{\overline{T}}\overline{W} _M}
\over {\overline{W} _M}} \cr
\partial_{z} {\rm ln} A &= 2\zeta (2^{-n}A{\rm e}^{2\sum ^n_1 \phi _{i}})^{1/2}
{\rm e}^{K_{M}/2}|W_{M}| \cr
\partial _{z}\phi _{i} &= -\zeta (2^{-n}A{\rm e}^{2\sum ^n_1 \phi _{i}})^{1/2}
{\rm e}^{K_{M}/2}|W_{M}| \cr}\eqn\5
$$
where $\zeta$ is either $+1$ or $-1$ and can change sign when and only when
$W$ vanishes.\refmark{\CGR, \CG}  The above coupled first order differential
 equations  can be viewed as ``square roots'' of the corresponding Einstein
and Euler-Lagrange equations; they provide special solutions
of equations of motion which  saturate the Bogomol'nyi bound \4\ .

The topological charge $|C|$ can be determined in the thin wall
approximation.  Then in the wall region ($z \sim z_0 = 0$, without loss of
generality) the matter
field $T$ is a quickly varying function, resembling   a step function centered
 at the wall,  while the metric $A(z)$ and
$\phi _{i}(z)$ fields  vary slowly.  With the choice  $A(0) = 1$ and the
boundary conditions $\phi _{i}(0) = (\phi _{i})_{0}$ ($i = 1,...,n$),  one
obtains\refmark{ \CGR ,\C }
$$
\sigma = |C| \equiv 2| (\zeta |W{\rm e}^{K/2}|)_{z=0^+}-(\zeta |W{\rm e}^{K/2}
|)_{z=0^-}| = 2(\alpha _1 \pm
\alpha _{2}) \eqn\6
$$
where $\alpha_{1,2}\equiv 2^{-{n \over 2}}\exp [\sum ^n_{1}
(\phi _{i})_{0}]{\rm e}^{{K_M} \over 2}
|W_M|_{1,2}$.  Here the subscript 1 [or 2] refers to the side of the wall
with the larger [or smaller] value of $\alpha$.  The $+$ and $-$ signs
correspond to a solution with $W_M$ crossing zero and $W_M \neq 0$
everywhere, respectively. Note that there are {\it no}  walls  corresponding
to $\alpha_1=\alpha_2=0$, {\it i.e.},  the superpotential $W_M$ has to have
non-zero value at least on one side of the wall.

The first two equations in \5\ describe the evolution of the
matter field $T = T(z)$ with $z$.  The first equation  is the ``geodesic''
equation \refmark\CGR\ for the complex $T$ field.  It is the same as for the
ordinary and dilatonic  supergravity walls.  The third and fourth equations in
\5\ for the conformal factor $A(z)$ and the real scalar
fields $\phi _{i}(z)$ imply:
$$
A(z){\rm e}^{2\phi _{i}(z)} = {\rm e}^{2(\phi _{i})_{0}}; \ \ \ \
i = 1,..., n. \eqn\7
$$
Here we have used  the boundary conditions $A(0) = 1$ and
$\phi_ {i}(0) = (\phi _{i})_{0}$.
Note, that this equation is true {\it everywhere} in the domain wall
background. If one  chooses to take  one of the fields, say,  $\phi_1$, to
be the dilaton field of the $4d$ string vacua   then  Eq.\7\
implies that the string
frame  metric ($A_s(z)\equiv A(z){\rm e}^{2\phi _{1}(z)}$) is {\it
flat} everywhere in the domain wall background.
For $n\ne 1$  the second equation \5\ for the matter field $T$ does
not decouple from  the conformal factor $A(z)$ and the scalar fields
$\phi_{i}(z)$
 and therefore, the evolution of $T(z)$ will  be
affected by the presence of $\phi _{i}(z)$.\refmark\FOOTK\
For a thin wall one sets
$\tilde{V}= (n-3){\rm e}^{K_M}|W_M|^2=const.$  outside the wall
and then the solutions  can be found explicitly.\refmark{\FOOTF ,\FOOTFF}\

One can classify  solutions into type $I$, $II$, and $III$ walls in a
 manner similar to  those of
 ordinary supergravity walls ($n=0$)\refmark{\CGR,\CG} and dilatonic walls
($n=1$).\refmark\C\
Here we concentrate\refmark\FOOTL\ on the type $I$ solutions  corresponding
to the case  where
 $\alpha _{1} \ne 0$  and
$\alpha _{2} = 0$, {\it i.e.}, on one side of the wall $W_M=0$.  The thin
wall  solution has the form:
$$
\sigma=2\alpha_1 \ ; \ \
 \left\{
\eqalign{A(z)
= [1 - (n - 1)\alpha_1 |z|]^{2 \over n-1},\ \phi_i={(\phi_i)}_0
-{1 \over n-1}{\rm ln}[1 - (n - 1)\alpha_1 |z|] ,\ \  & z<0 \cr
A(z) = 1,\ \ \ \ \ \ \ \ \ \ \ \ \ \ \ \ \ \ \ \ \ \ \ \ \ \ \ \ \ \ \ \ \ \
\phi_i={(\phi_i)}_0, \
 \ \ \ \ \ \ \ \ \ \ \ \ \ \ \ \
  &z >0 . }
\right.
\eqn\8
$$

On one side of the wall ($z>0$)
the space-time is flat, and on the other side ($z<0$) $A(z)$ vanishes
at the finite coordinate
distance $|z|_{sing} = {1 \over (n - 1)\alpha_1}$ from the wall.
Both the scalar fields $\phi _{i}(z)$ and the curvature invariants
blow up in this region.
\refmark\FOOTH\  For $n = 2$, $R = 0$ but $R_{\mu \nu}R^{\mu \nu}
= \infty$ at the singularity. For $n > 2$, not only
$R_{\mu \nu}R^{\mu \nu} = \infty$ but also $R = 3\cdot 2(2 - n)
\alpha ^2[1 - (n - 1)\alpha_1|z|]^{- {2n \over n-1}}$
blows-up  at $|z|_{sing}$ and the space-time becomes more
singular  as the number $n$ of dilatons  $\phi _{i}(z)$ increases.  Clearly,
$|z|_{sing}$ is a  finite proper distance  $d = \int \sqrt
{A(z)}dz = {1 \over n\alpha_1}
< \infty$  as well as  within a finite affine
parameter from the wall.\refmark\FOOTI\  Thus, the  singularity is naked.
\refmark\FOOTM

One can trace the origin of the naked singularity
to the nature of the  stress-energy tensor on the side of the wall
with varying dilaton fields. The stress-energy  is diagonal with non-zero
components:
$T_{tt}=-T_{xx}=-T_{yy}=(2n-3)\alpha_1^2[1-(n-1)\alpha_1|z|]^{-2}$ and
$T_{zz}=3\alpha_1^2[1-(n-1)\alpha_1|z|]^{-2}$.    Thus, it
 satisfies the weak energy
condition $T_{tt}\ge 0$  for $n\ge 2$ and the dominant energy condition
$T_{tt}\ge |T_{ii}|$  $(i=x,y,z)$ for $n\ge 3$; however it violates the
strong energy condition  $T^t_t-{1\over 2}\sum_{i=t,x,y,z} T^i_i\ge 0$
for any $n\ge 2$.
Our supersymmetric  solutions  are thus in agreement with the theorem
\refmark \DK\
that static planar solutions   are singular when the stress-energy satisfies
the
weak energy condition  $T_{tt} \ge 0$.  On the other hand, such new
supersymmetric vacua    violate the
strong  energy condition,  which is in sharp contrast with
the   Minkowski and anti-deSitter
space-times, {\it i.e.}, unique supersymmetric vacua without matter sources.
Only the latter ones   were assumed to be the asymptotic space-times
in the proof  of  ``no solitons without horizons'' in the
ungauged extended supergravity.\refmark\GB\

We found  new   {\it supersymmetric} vacuum domain walls  within a class
of $4d$, $N=1$ supergravity models, where along with the matter
field forming the wall there are  $n\ge 2$ dilatons, each of them respecting
$SU(1,1)$ symmetry in their sub-sector.
Such walls  interpolate
between  (supersymmetric) Minkowski  vacuum  and a  new
class of supersymmetric vacua  associated  with a non-trivial
source of the dilatons whose stress-energy tensor violates the
strong energy condition.
Consequently, while on one side of the wall the space-time is flat, on the
other side there is  a naked singularity.
Such walls correspond to the first example of
classical supersymmetric solitons  with naked singularities, and thus
violate  the (strong) cosmic censorship conjecture within a supersymmetric
theory.

We would like to thank G. Gibbons, G. Horowitz and R. Kallosh
for useful discussions. The work is supported by  U.S. DOE  Grant No.
DOE-EY-76-02-3071.
\endpage

\refout

\end